\documentclass[
reprint,
a4paper,
secnumarabic,
showpacs,showkeys,preprintnumbers,
nofootinbib,
 amsmath,amssymb,
 aps,
]{revtex4-1}

\usepackage{graphicx}
\usepackage{dcolumn}
\usepackage{bm}

\begin{document}

\title{MULTI-SPACE STRUCTURE OF THE UNIVERSE}

\author{A.V.~Novikov-Borodin}
\affiliation{Institute for Nuclear Research of RAS, 60-th October Anniversary prospect~7a, 117312 Moscow, Russia}
\begin{abstract} 
If our universe has appeared in a result of Big Bang or something like this, whether we have reasons to deny an existence of other universes appearing by the same or similar way? An objection that there is no anything like it, is doubtful, because nobody knows: what could we observe in this case? A model of a multi-space Universe with mutual coupling of spaces is being proposed and investigated. 

\begin{description}
\item[Keywords]
Multi-Space model, General relativity, Quantum theories, Unified Field theories, Set theory.
\item[PACS numbers]
98.80.Bp, 04.20.Cv, 12.10.-g, 02.10.Ab.
\end{description}
\end{abstract}
%
\maketitle
%
%

\section{\bf Off-Site Spaces}
\label{sec:OSS}

\begin{figure}[t]
	\begin{center}
		\includegraphics*[angle=270,width=85mm]{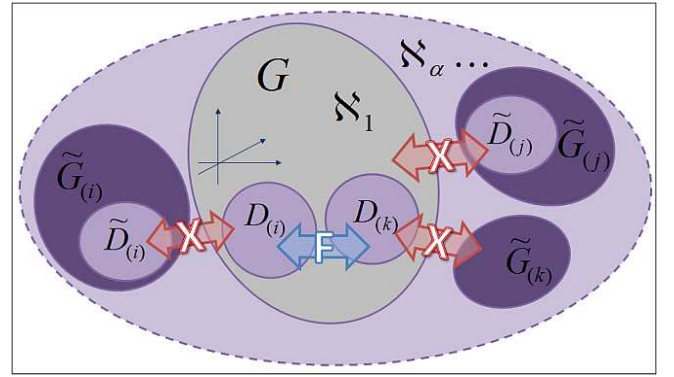}
	\caption{Coupling of Spaces} 
	\label{fig1}
	\end{center}
\end{figure}

Asserting an appearance of our universe, nobody has reasons to deny an existence of other `universes'. This supposition would be quite speculative without introducing interconnections between universes. In the general case, without limiting the kind of interconnections and without fixing the structure of `universes', one may consider that a spacetime $G$ of our universe is coupled by some interconnections $X$ with off-site spaces $\widetilde G$ of other universes on some shared regions: $G\supseteq D {\rightleftharpoons} \widetilde D \subseteq \widetilde G$. This general case is shown on Figure~\ref{fig1} for a space $\widetilde G_{(i)}$ together with two particular cases: $G\supseteq D {\rightleftharpoons} \widetilde D\equiv \widetilde G$ for {\it included} off-site space $\widetilde G_{(k)}$ and $G\equiv D {\rightleftharpoons} \widetilde D \subseteq \widetilde G$ for {\it containing} one $\widetilde G_{(j)}$.

When the coupling $X$ results in some fields $\Lambda_m$ on $D$ and $\widetilde \Lambda_m$ on $\widetilde D$, the observers in $G$ and $\widetilde G$ will identify this coupling with physical systems with actions: $S_m=\int_D \Lambda_m d\Omega$ and $\widetilde S_m =\int_{\tilde D} \widetilde \Lambda_m d\widetilde \Omega$. Here, in $G$: $d\Omega=\sqrt{-g}dx$ is the elementary volume, $g$ the determinant of metric tensor $g^{ik}$, $x$: $(x^0,...,x^n)$ coordinates. In particular, when $D\equiv G$: $S_M=\int_G \Lambda_M d\Omega$. Similar notions may be introduced in $\widetilde G$. Thus, the coupling $X$ between spaces and interactions $F$ inside a spacetime $G$ (see Figure~\ref{fig1}) are parts of general interconnections of the separated physical system.

In general relativity (GR) a gravitational field is being described with help of the action $S_g$. Its variation by the elements of the metric tensor $g^{ik}$ is: $\delta S_g = -\delta \int_G R d\Omega$, where $R=g^{ik}R_{ik}$ is a curvature in $G$, $R_{ik}$ the Ricci tensor. Thus, the variational equation for coupled spaces is: 
\begin{equation}
\delta \int_G (-R+\Lambda_M) d\Omega+ \sum \delta \int_D \Lambda_m d\Omega=0.  
\label{VE}
\end{equation} 

Approximating: $\Lambda_M\approx -2\Lambda$, where $\Lambda$ is a cosmological constant, one comes to Einstein's field equations with the $\Lambda$-term:  
\begin{equation}
R_{ik}-\frac{1}{2}g_{ik}R= \frac{8\pi {\cal G}}{c^4} T_{ik}+\Lambda g_{ik},  
\label{EFE}
\end{equation} 
where ${\cal G}$ is a gravitational constant, $T_{ik}$ an energy-momentum tensor corresponding to $\Lambda_m$, $c$ a speed of light. The $\Lambda$-term $\Lambda g_{ik}$ is an approximation to the energy-momentum tensor $T^M_{ik}$ from $\Lambda_M$, so $ T^M_{ik}\approx \hat \varepsilon g_{ik} $, where $\hat \varepsilon={c^4 \Lambda}/({8\pi {\cal G}})$. 

When $g^{ik} = {\rm diag}(1,-1,-1,-1)$, comparing $ T^{ik}_M \approx {\rm diag}(\hat \varepsilon,-\hat \varepsilon,-\hat \varepsilon,-\hat \varepsilon)$ with the energy-momentum tensor of a perfect fluid: $ T^{ik}_{PF}= {\rm diag}(\varepsilon,p,p,p)$, one will find that an energy density of the containing off-site space in $G$ is: $\varepsilon = \hat \varepsilon >0$ and its `pressure': $ p=- \hat \varepsilon <0$. It corresponds to the model of {\it dark energy} (DE) in modern cosmology. The negative `pressure' seems unphysical, but the trace $ T^M{^i_{i}}=4\varepsilon \geq 0$. It agrees with the virial theorem, so, most probably, the properties of dark energy seem unnatural only in the rough model used by the observer. 
 
In fact by definition, the physical objects of included spaces will always be observered inside the shared region $D\subset G$. An observer interprets this visual `capturing' of off-site objects in scale of elementary particles as a confinement and in cosmological scale he needs to suppose an existence of some invisible \emph{dark matter} (DM) attracting visible physical objects. 

To illustrate above statements, let's tentatively consider the stable ($\tau=\tilde \tau$) coupling $X$: $ r(\tilde r) = \arctan \tilde r$ (or $ \tilde r(r) = \tan r$) between two spaces $G$: $(x^0{,}\ldots{,}x^{n})=(\tau, \mathbf{r} )$, $r=| \mathbf{r} |$ and $\widetilde G$: $(\tilde x^0{,}\ldots{,}\tilde x^{m})=(\tilde \tau, \tilde {\mathbf{r}} )$, $\tilde r=|\tilde {\mathbf{r}} |$. This coupling reflects the $m$-dimensional spatial subspace of included space $\widetilde G$ into the $n$-dimensional  hypersphere of radius $r=\frac{\pi}{2}$ of containing space $G$. 

\begin{figure*}[t]
	\begin{center}
		\includegraphics*[angle=270,width=175mm]{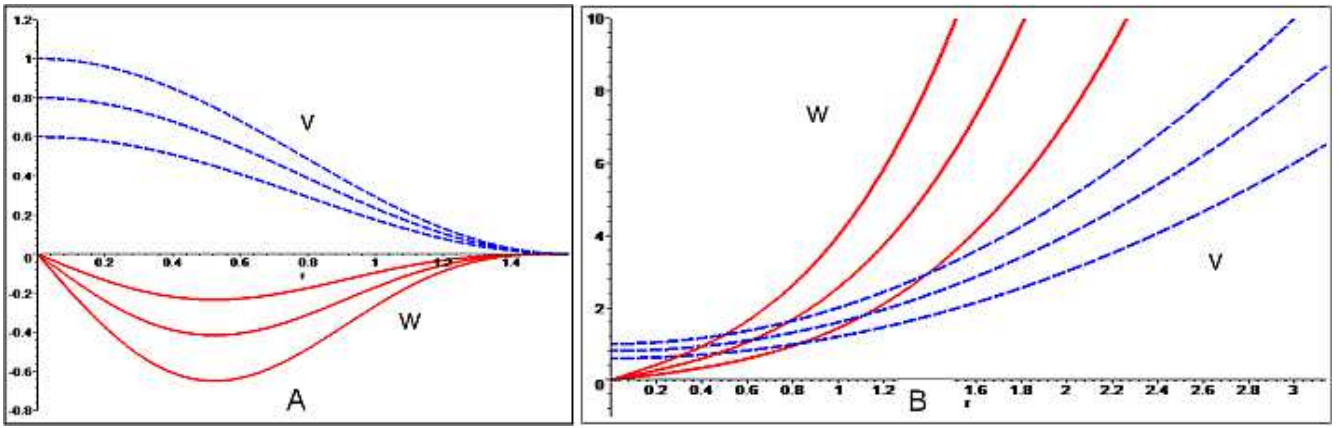}
	\caption{Observations of objects from included (A) and containing (B) off-site spaces}
	\label{fig2}
	\end{center}
\end{figure*}

For this coupling, any physical object moving with the constant radial velocity $\tilde v = \dot {\tilde r }= d \tilde r/d \tau $ in $\widetilde G$ (so, its radial acceleration $\tilde w = \ddot {\tilde r }= 0$) will be seen in $G$ as moving with the visual velocity $v= \dot r(\tilde r) =\tilde v / \tilde r'(r) = \tilde v \cos^2 r$, where $ \tilde r'(r) = d\tilde r(r)/dr $, and with the visual acceleration $w= - \tilde v^2 \tilde r''(r) /\tilde r'(r)^3 = - \tilde v^2 \sin 2r \cos^2 r$. Thus, one observes the physical objects of included space like moving with visual centripetal acceleration and their visual radial velocity $v \rightarrow 0$ with $r \rightarrow \frac{\pi}{2} $ (see Figure~\ref{fig2}A). For other coupling $X$: $ r(\tilde r) = 1-e^{-\tilde r}$ one will come to similar conclusions, but the visual acceleration and velocity will be: $w=-\tilde v^2 (1-r)$ and $v=\tilde v (1-r)$. 

The coupling $X$: $\tilde r(r) = \arctan r$, reversed to just considered, reflects the $n$-dimensional spatial subspace of included space $G$ into the $m$-dimensional hypersphere of radius $\tilde r= \frac{\pi}{2} $ of containing space $\widetilde G$. Now any physical object moving with the constant radial velocity $\tilde v$ in $\widetilde G$ will be seen in $G$ as moving with the visual velocity $v= \tilde v (1+r^2) $ and with the visual acceleration $w= 2\tilde v^2 r (1+r^2)$. Thus, one observes the physical objects of containing space like expanding with visual centrifugal acceleration (see Figure~\ref{fig2}B). For the coupling $X$: $ \tilde r(r) = 1-e^{- r}$: $v=\tilde v e^{r}$ and $w= {\tilde v}^2 e^{2r}$, the physical objects of containing spaces are looking expanding exponentially exactly as in the de Sitter model of the expanding universe. 

In the general case, one can formally introduce the coordinate transformations $X$ between spaces $G$: $(x^0{,}\ldots{,}x^{n})$ and $\widetilde G$: $(\tilde x^0{,}\ldots{,}\tilde x^{m})$ as: $x^i=x^i(\tilde x)$, $i=0..n$ or $\tilde x^j=\tilde x^j(x)$, $j=0..m$, which at any fixed point are:
\begin{equation}
d\tilde x^j= X^{j}_{i}dx^i,\quad dx^i= X^{i}_{j} d\tilde x^j, 
\label{X}
\end{equation}
where a summation on repeating indexes is meant, $X^{j}_{i}= {\partial \tilde x^j}/{\partial x^i} $ are $(m+1)\times(n+1)$ matices and $X^{i}_{j}={\partial x^i}/{\partial \tilde x^j} $ are $(n+1)\times(m+1)$ ones. Formal expressions for vector and tensor transformations: $A_i=X^{j}_{i} \widetilde A_j$, $T_{ik}=X^{j}_{i} X^{l}_{k} \widetilde T_{jl}$, etc. look similar to curvilinear transformations of coordinates. 

According to Eq.(\ref{X}), an interval in $\widetilde G$ seen by an observer from $G$ is: $d\tilde s^2= \tilde g_{jl} d\tilde x^j d\tilde x^l= \tilde g_{jl} X^{j}_{i} X^{l}_{k} dx^i dx^k = \hat g_{ik} dx^i dx^k$. Generally, the visible metric tensor $\hat g_{ik}$ differs from $g_{ik}$ (see Section~\ref{sec:PoC} for details): 
\begin{equation}
g_{ik} \neq \hat g_{ik} = \tilde g_{jl} X^{j}_{i} X^{l}_{k}, 
\label{g} 
\end{equation}
so a visible motion of off-site objects in $D$ differs from expected by the observer. 

In GR, one can consider \emph{any} curvilinear transformations, but not all of them may be `physically realized by real bodies'\cite{LL88}. However, curvilinear transformations are usually considered in the same space, while the coupling interconnects different spaces, so there are no such limitations for objects in off-site spaces. 

The visual properties of off-site objects depend on the visible metric tensor $\hat g_{ik}$ from Eq.(\ref{g}). One can adopt the analysis of L.Landau and E.Lifschitz\cite{LL88} of curvilinear transformations for the coupling with $\hat g_{ik}$ (see also A.Novikov-Borodin\cite{NB08} for details). One may separate the areas in $D$ where the sign of $\hat g_{00}$ is positive or negative and where the principal values of $\hat g_{ik}$ are $(+,-,-,-)$ or not: 

$\bullet$ The \emph{timelike area} [$\hat g_{00}>0$, $(+,-,-,-)$]: The off-site physical objects are identified with an `ordinary matter' in $G$, but their motion in $D$ is unexpected, because it is defined by the visual metric tensor $\hat g_{ik}\neq g_{ik}$.  

$\bullet$ The \emph{transitive area} [$\hat g_{00}<0$, $(+,-,-,-)$]: \emph{``The non-fulfillment of the condition ($\hat g_{00}>0$) would mean only, that the corresponding frame of references cannot be realized by real bodies; thus if the condition on principal values is carried out, it is possible to achieve ($\hat g_{00}$) to become positive by an appropriate transformation of coordinates''} (\cite{LL88}). The off-site objects cannot be identified with an `ordinary matter' in $G$, but they `act like real bodies'. In cosmology \emph{``some invisible distributed in space substance, strange `amorphous' media interacting gravitationally with identified visible objects''} is introduced as \emph{dark matter}. 

$\bullet$ The \emph{spacelike area} [principal values $\neq (+,-,-,-)$]: \emph{``The tensor ($\hat g_{ik}$) cannot correspond to any real gravitational field at all, i.e. the metrics of the real space-time''}\cite{LL88}. There is no possibility to identify both off-site objects and their influence with `real bodies', so off-site objects will be invisible and possess unphysical characteristics (such as, for example, the gravitational repulsion or negative pressure). In cosmology such nonphysical characteristics are inherent to {\it dark energy}. 

$\bullet$ The \emph{unreachable area}: The physical objects from [$\widetilde G\setminus \widetilde D$] are out of the horizon of events of a spacetime $G$, but their influence to $\tilde g_{jl}$ on $\widetilde D$ may be registered. 

The further analysis of off-site objects needs more detailed consideration of a process of coupling. 

\section{\bf Process of Coupling}
\label{sec:PoC}

Let's some functions $\Lambda_0 (x)$ and $\widetilde \Lambda_0 (\tilde x)$ are defined on $D_0\subset G$ and $\widetilde D_0\subset \widetilde G$ before coupling. During the coupling $\Lambda_0 (x)$ is transformed from $G$ by some transformation $\widetilde {\rm K} \Lambda_0$ to $\widetilde G$ and then is combined with $\widetilde \Lambda_0 (\tilde x)$ according to some functional $\widetilde {\rm F}$ in $\widetilde G$, so the result is: $\widetilde \Lambda_1 (\tilde x) = \widetilde {\rm F} ( \widetilde {\rm K} \Lambda_0, \widetilde \Lambda_0)$, where $\tilde x \in\widetilde D_1\subset \widetilde G$. The similar process with $\widetilde \Lambda_0 (\tilde x)$ results in: $\Lambda_1 (x)={\rm F} ({\rm K} \widetilde \Lambda_0, \Lambda_0)$, $x \in D_1\subset G$. These steps are sequentially repeated for functions $\Lambda_1$ and $\widetilde \Lambda_1$, then for $\Lambda_2$ and $\widetilde \Lambda_2$, etc., so the coupling $X$ is a process of multiple iterations: 
\begin{equation} 
X:
\left[\begin{array}{ll} 
\Lambda_{i+1} = {\rm F} ({\rm K}\widetilde \Lambda_i, \Lambda_i) \\ 
\widetilde \Lambda_{i+1} = \widetilde {\rm F} (\widetilde {\rm K} \Lambda_i, \widetilde \Lambda_i) 
\end{array}\right., \quad i=0..\infty.  
\label{IP} 
\end{equation} 

Functions $\Lambda_0$ and $\widetilde \Lambda_0$ initiate two branches: $\Lambda^{a}_{2i}$ ($ \Lambda^{a}_{0}=\Lambda_0$) appear on $D^{a}_{2i}\subset G$ at odd steps of iterating, while $ \Lambda^{b}_{2i+1} $ ($ \widetilde \Lambda^{b}_{0} = \widetilde \Lambda_{0}$) appear on $D^{b}_{2i+1}$ at even steps, so branches alternate in $G$: $\Lambda^{a}_0 (D_0) \to \Lambda^{b}_1 (D_1) \to \Lambda^{a}_2 (D_2) \to \cdots$. 

When $\Lambda^{a,b}_{i} \to \Lambda^{a,b}$ with $i\to \infty$, the coupling looks stable in time $\tau$ on spatial regions $V^{a,b}$ ($D$: $\tau\times V$). If stable states $\Lambda^{a,b}$ appear periodically on $V^{a,b}$ with the frequency $\omega$, one can use Fourier series for a spectral decomposition of the resulted function $\Lambda (\tau, V) $:
\begin{equation} 
\Lambda (\tau, V) = \sum^{\infty}_{k=-\infty} c_k (V) e^{\imath k \omega \tau}, \quad 
c_k = \bar c_{-k}= \langle \Lambda e^{- \imath k \omega \tau} \rangle_T, 
\label{Lk} 
\end{equation} 
where $\langle f \rangle_T=\frac{1}{T}\int^{T/2}_{-T/2} f(\tau) d\tau$, $T=\frac{2\pi}{\omega}$. 

In case of coupling of three spaces: $G_1 \rightleftharpoons \widetilde G_2 \rightleftharpoons \widetilde G_3$, four iteration processes will be initiated: three ($C^2_3=\frac{3!}{2!(3-2)!}$) between each two spaces and one ($C^3_3$) between three spaces with different `spinning': $G_1 \to \widetilde G_2 \to \widetilde G_3$ and $G_1 \to \widetilde G_3 \to \widetilde G_2$. If stable states $\Lambda^{a,b,c}$ exist, an observer will see their appearance in $G$ with frequencies corresponding to different iteration processes. If $\omega_2 = \omega$ characterizes processes between each two spaces, so $\omega_3= \frac{2}{3} \omega$ will correspond to processes between three spaces. 

For a system of $N$ coupled spaces, there will be $C^n_N$ processes between $n$ spaces with characteristic frequencies $\omega_n = \frac{2}{n} \omega $, so $\sum^N_{n=2}C^n_N=2^N-N-1$ iteration processes each with two branches will be initiated. One can generalize Eq.(\ref{Lk}) for a stable state of $N$ coupled spaces: 
\begin{equation} 
\Lambda_N (\tau,V) = \sum^{\infty}_{k=-\infty} \sum^{N}_{n=2} c_{nk} (V) e^{\imath k \omega \tau/n}. 
\label{LN} 
\end{equation} 

Representations~(\ref{Lk},\ref{LN}) are also followed from probabilistic approach. Indeed, considering $\psi(\tau)$ as a quasi-stationary process on $V$ one can use a spectral expansion: 
\begin{equation} 
\psi (\tau, V) = \sum^{\infty}_{l=-\infty} \psi_l (V) e^{\imath \omega_l \tau} + p(\tau),
\label{SE} 
\end{equation} 
where $\langle\psi e^{-\imath \omega \tau} \rangle_T= \psi_l =\bar \psi_{-l} $ for $\omega = \omega_l$, $l=0,\pm 1..$, and for an aperiodic component: $\langle p(\tau) e^{-\imath \omega \tau} \rangle_T = 0$. Here, $\langle f \rangle_T = \frac{1}{2T}\int^T_{-T}  f(\tau) d\tau$ with ${T\rightarrow \infty} $. Identifying $\Lambda_N$ with $\psi$, $\omega_l$ with $\frac{k}{n} \omega$ and $\psi_l$ with $c_{nk}$, one comes from Eq.(\ref{SE}) to Eqs.(\ref{LN}) or (\ref{Lk}). 

An observer identifies the coupling on $D$ or $V$ with the physical system, so integrals:
\begin{equation} 
q(D) = \int_D \Lambda d\Omega, \quad q(V) = \int_V \Lambda dV  
\label{qD} 
\end{equation} 
are, in fact, an action $S_m =- q(D)$ and a Lagrangian $L = -c q(V)$ of this system. The integrand $ \Lambda $ is understood as a `distribution' $q(x)$ of the property $q$ on $D$ or $V$. One can represent $\Lambda = q(x) + A^i B_i + C^{ik}D_{ik} + \cdots$ to consider vector or tensor properties of the physical system. 

By default, spaces are understood as continuums or manifolds, so any continuous function $\widetilde \Lambda (\tilde x)$ from an off-site space $\widetilde G$, in fact, by definition, is NOT seen as continuous in $G$, but as a set of discontinuities, also as $\Lambda (x)$ in $\widetilde G$. Thus, generally, integrands $\Lambda$ in Eq.(\ref{qD}) are discontinuous and even non-integrable, so an action and a Lagrangian are not defined. We need to introduce a notion of {\it integrally generalized} functions to continue the analysis of the spaces' coupling. 

\section{\bf Integrally Generalized Functions}
\label{sec:IGF}

When an integral property $q=q(D)$ or $q=q(V)$ of an off-site physical object is identified (usually by interactions with other objects), integrals in Eq.(\ref{qD}) have definite values while the integrand $\Lambda$ is not integrable. It may be considered as a definition of the {\it integrally generalized (IG) function} $q_\xi(D)$ on $D$:
\begin{equation} 
\int_D q_\xi (D) d\Omega = \int_D q\delta_\xi (D) d\Omega = q, 
\label{qDD} 
\end{equation} 
where $ \delta_\xi (D)$ is an IG delta-function. For $q = 1$ one has: 
\begin{equation} 
\int_{D^+} \delta_\xi (D) d\Omega = 
\left[\begin{array}{lr} 
1, & \textrm{ when } D \subset D^+ \\ 
0, &\textrm{ when } D \not \subset D^+ \\ 
\textrm{\it undefined}&\textrm{ in other cases.}  
\end{array}\right.
\label{dD} 
\end{equation} 
One can see that a Dirac delta-function is a passage to the limit of an IG delta-function: $\delta_\xi(D) \to \delta (x_0)$ and $q_\xi(D) \to q\delta (x_0)$ with $D\to x_0 \in D$. 

One can approximate both IG $q_\xi (D) $ and Dirac $q\delta(x_0)$ delta-functions as: 
\begin{equation} 
q(x)= \sum^N_{n=1} q_n \delta ( x_n), \quad
\left[
\begin{array}{l} 
\sum_n q_n = q \\
\forall n: x_n \in D 
\end{array}\right.,
\label{qxD} 
\end{equation} 
because $q(x) \to q_\xi (D) $ with $N \to \infty$ and $q(x)\to q\delta(x_0)$ with $D \to x_0 \in D $. 

In particular, for $x$: $(\tau,\mathbf{r})$ and $D$: $\tau \times V$, a function: 
\begin{equation} 
q(\tau, \mathbf{r}) = \sum^N_{n=1} q_n (\tau) \delta (\mathbf{r}_n), \quad 
\left[
\begin{array}{l} 
\forall \tau: \sum_n q_n (\tau) = q \\
\forall n: \mathbf{r}_n \in V 
\end{array}\right.
\label{qxV} 
\end{equation} 
$\forall \tau$: $q(\tau, \mathbf{r} ) \to q_\xi (V) $ with $N \to \infty$ and $\forall \tau$: $q(\tau,\mathbf{r} )\to q\delta(\mathbf{r} _0) $ with $V \to \mathbf{r} _0 \in V $. 

In an IG formalizm, real physical bodies are not only a combination of points: $\sum_n q_n \delta(x-x_n)$ or $\int q(y) \delta(x-y) d\Omega(y)$, but are a combination of IG functions: 
\begin{equation} 
q_\xi (D) \leq \sum^N_{n=1} q_n \delta_\xi (D_n), \quad \sum^N_{n=1} q_n =q. 
\label{IGx} 
\end{equation} 
There is an equality here only if regions $D_n$ do not intersect with each other. 

In fact, instead of Dirac delta-functions $q \delta (x_0)$ or $q \delta ({\mathbf{r}}_0)$, there is introduced a new {\it elementary entity} -- the IG delta-functions $q\delta_\xi (D)$ or $q\delta_\xi (V)$, which have not a point character, but are an integral property of some regions $D$ or $V$. 

\section{\bf IG Fields}
\label{sec:IGFF}

Let's consider a second-order differential equation: 
\begin{equation} 
\nabla^2 \phi_p(x)= {\rm div}\,{\rm grad}\, \phi_p(x)= q \delta (x_p),  
\label{GP}
\end{equation}
which puts some functions $\phi_p(x)$ in correspondence to a Dirac delta-function $q \delta (x_p)$. An operator $\nabla^2 = {\rm div}\,{\rm grad}$ in curvilinear coordinates is: $\nabla^2 = \frac{1}{\sqrt{|g|}} \frac{\partial}{\partial x^i} \left( \sqrt{|g|} g^{ik}\frac{\partial}{\partial x^k} \right)$, which for $x$:$(\tau, \mathbf{r})$ gives a $3D$-wave operator: $\nabla^2_{\tau \mathbf{r}} = \frac{\partial^2}{\partial \tau^2} -\nabla^2_{\mathbf{r}} $, for $x$: $(\tau,x)$: $\nabla^2_{\tau x} = \frac{\partial^2 }{\partial \tau ^2}- \frac{\partial^2 }{\partial x^2}$ and in $1D$-case: $\nabla^2_{ x} = \frac{d^2 }{d x^2}$. 

Functions $\phi_p(x)$ are fundamental solutions of Eq.(\ref{GP}) (see, for example, V.S.Vladimirov\cite{Vlad81}). With $\nabla^2 = \nabla^2_{\mathbf{r}}$, they correspond to a gravitational or electrostatic potential from a point-like source in $x_p$ when $q$ is a mass $m$ or a charge $e$. It is a wave equation when $\nabla^2 = \nabla^2_{\tau \mathbf{r}}$, or Maxwell's equations for electromagnetic fields with a vector potential: $\phi_i=A_i$, $i=0..3$ produced by `currents' $q_i=\alpha j_i$ (see comments to Eq.(\ref{qD})), or some tensor fields $\phi_{ik}$, $i,k=0..3$ from $q_{ik}=\alpha T_{ik}$, etc. Thus, Eq.(\ref{GP}) is a {\it classical} field equation of from a point-like {\it source} $q \delta (x - x_p)$. 

In an IG formalizm, physical objects are described by IG functions, so an {\it IG field equation} is: 
\begin{equation} 
\nabla^2 \phi_\xi (x)= q_\xi (D), \quad \int_{\partial D} \nabla \phi \,\mathbf{n} dS = q, 
\label{FE} 
\end{equation} 
where an edge condition comes from Eq.(\ref{qDD}). 
Taking into account approximations~(\ref{qxD},\ref{qxV}) for $q_\xi (D)$, one comes to the equation: 
\begin{equation} 
\nabla^2 \phi(x)= \sum^N_{n=1} q_n \delta ( x_n), \quad
\left[
\begin{array}{l} 
\sum_n q_n = q \\
\forall x_n \in D 
\end{array}\right.,
\label{FEx} 
\end{equation} 
which converges to Eq.(\ref{FE}) when $N\to \infty$ and to Eq.(\ref{GP}) when $D\to x_p \in D$. 

\begin{figure}[t]
	\begin{center}
		\includegraphics*[angle=270,width=85mm]{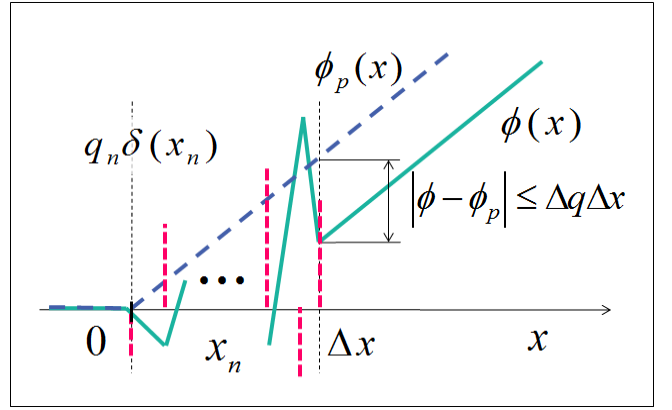}
	\caption{Uncertainty of IG fields} 
	\label{fig3}
	\end{center}
\end{figure}

In a $1D$-case: $\nabla^2=\nabla^2_{x} = \frac{d^2}{d x^2}$, and Eq.(\ref{FEx}) has solutions: 
\begin{equation} 
\frac{d^2 }{d x^2}\phi (x) = \sum^N_{n=1} q_n \delta(x_n): \quad \phi(x) = - x \sum^{N}_{n=1} q_n \theta(x_n), 
\label{1D} 
\end{equation} 
where $\sum_n q_n = q$, $\forall x_n \in D$: $[0,\Delta x ]$, $\theta (x)$ is a step function and, for definiteness, the edge condition $\phi(x<0)=0$ is applied. The fields $\phi(x)$ depend on distributions $\{q_n\}$ and $\{x_n\}$. If a range of $q_n$ on $V$ is: $\Delta q = q^{max} - q^{min}$, so: $ |\phi(x)-\phi_p (x)| \leq \Delta q \Delta x$ (see Figure~\ref{fig3}). With $N\to \infty$ this deviation is an {\it uncertainty} in determining and measuring the fields $\phi_\xi (x)$ from IG sources $q_\xi(D)$, so $\phi_\Delta =| \phi_\xi -\phi_p | $ : 
\begin{equation} 
\Delta q \Delta x \leq |\phi_\Delta|.  
\label{UC} 
\end{equation} 

Generally, IG fields are nondeterministic and have an uncertainty~(\ref{UC}), but one may find such distributions $\{q_n\}$ on $\{x_n\}$, for which a deviation $\phi_\Delta =0$ when $x\not\in [0,\Delta x]$, i.e. for some class of sources, outside the region $D$ the IG fields coincide with fields from corresponding point-like source. For example, fields from sources $2q\delta(\frac{1}{2}\Delta x)- q\delta (\Delta x)$ coincide with fields from $q \delta (0)$ on $x\not \in [0, \Delta x]$. 

\section{\bf IG Systems}
\label{sec:IGS}

We will show below, that the fields $\phi_\xi (\tau, {\mathbf r})$ from some IG sources $q_\xi (V)$ in Eq.(\ref{FE}) may coincide outside the region $V$ with the corresponding field $\phi_p (\tau, {\mathbf r} )$, produced by the point-like source $q \delta({\mathbf r}_p)$, ${\mathbf r}_p \in V$, so for $\forall \tau$ and ${\mathbf r}\not \in V$ a condition:  
\begin{equation} 
\phi_\Delta ({\mathbf r}\not \in V) = \left | \phi_\xi (\tau, {\mathbf r}) - \phi_p (\tau, {\mathbf r} )\right | _{|{\mathbf r}\not \in V}  = 0 
\label{QC} 
\end{equation} 
is satisfied. In fact, it is a condition of {\it quantization} of the {\it IG system}. 

Let's consider Eq.(\ref{FEx}) for a $1D$-wave operator $ \nabla^2= \nabla^2_{\tau x}$: 
\begin{equation} 
\frac{\partial^2 \phi (\tau,x)}{\partial \tau ^2}- \frac{\partial^2 \phi (\tau,x)}{\partial x^2}= \sum^N_{n=1} q_n (\tau) \delta (x_n).  
\label{1W} 
\end{equation} 

Representing in accordance with Eqs.(\ref{Lk},\ref{LN}) the IG sources by harmonic oscillators: $q(\tau, x_n) = q \delta(x_n) e^{\imath \omega \tau}$, from Eq.(\ref{1W}) one can get fields: 
\begin{equation} 
\phi(\tau, x) = \sum^N_{n=1} \phi_n(\tau, x) = \sum^N_{n=1} \alpha q_n e^{\imath \omega (\tau - |x-x_n|)}, 
\label{1F} 
\end{equation} 
which are a superposition of waves spreading from each point $x_n$ in both directions. 

\begin{figure}[t]
	\begin{center}
		\includegraphics*[angle=270,width=85mm]{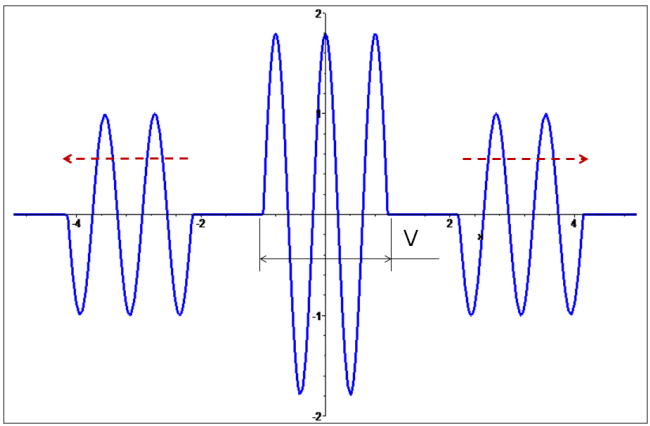}
	\caption{Steady-states of IG systems} 
	\label{fig4}
	\end{center}
\end{figure}

The condition~(\ref{QC}): $\phi_\xi (x\not\in V) = 0$ may be satisfied, for example, for a system of two synchronous harmonic sources: $q e^{\imath \omega \tau} \delta(0)$ and $q e^{\imath \omega \tau} \delta(\Delta x)$, if the distance between them is {\it quantized}: $\Delta x = \lambda (n-\frac{1}{2})$, $\lambda = \frac{2\pi}{\omega}$, $n=1,2..$. Indeed, for $\forall \tau$ the waves from these sources are in antiphase in $x\not \in V$: $[0,\Delta x]$, so compensate each other: $\phi (x\not\in V) = 0$. There are the standing waves: $\phi(x)= 2 \alpha q \sin(\omega x) e^{\imath\omega \tau}$ between sources inside the region $V$ (see Figure~\ref{fig4}). Note, that if the sources $q(0)$ and $-q(\Delta x)$ are in antiphase, the conditions of quantization are: $ \Delta x = \lambda n$, $n=1,2..$. 

This IG system does not emit or absorb waves only in steady-states $S_n$ corresponding to discrete sizes $\Delta x_n $ and passes from one state to another: $S_n (\Delta x_n ) \to S_k (\Delta x_k )$ with emission ($n>k$) or absorption ($n<k$) the wave quanta with a `length' $\lambda (n-k)$. Emitted quanta are shown on Figure~\ref{fig4} with arrows. 

The simplified mechanism of quantization of this IG system is the following. Two synchronous harmonic sources $q(0)$ and $q(\Delta x)$: $q(x) = q \delta (x) \sin({\omega \tau})$ act to each other by forces defined by their fields-potentials (Eq.(\ref{1F})): $F_1=-F_2  = - q(\tau) \frac{\partial}{\partial x }\phi(\Delta x)$, which may be represented as: 
$$\begin{array}{ll} 
F (\tau,\Delta x)& = \pm \alpha q^2 \omega \sin(\omega \tau ) \cos\left[{\omega (\tau - \Delta x)}\right] =\\ 
 &=\pm \frac{1}{2} \alpha q^2 \omega \left\{ \sin\left[\omega(2\tau - \Delta x )\right] + \sin (\omega\Delta x)\right\}.
\end{array}$$ 

This force averaged in time is: 
$$
\langle F (\tau, \Delta x)\rangle_\tau= \pm \frac{1}{2} \alpha q^2 \omega \sin (\omega\Delta x). 
$$

The condition of stable equilibriums for repulsive forces is: $ \Delta x_n = \lambda (n-\frac{1}{2})$, $n=1,2..$, while for attractive forces: $\Delta x_n = \lambda n$, $n=1,2..$. One may compare these results with the previous ones. 

The dynamical equation of the IG system: 
\begin{equation} 
m\frac{d^2}{d\tau^2} \Delta x (\tau) = \pm F(\tau,\Delta x) 
\label{DE} 
\end{equation} 
describes the non-linear oscillations of harmonic sources with inertial masses $m$ near the points of their equilibriums. In contrast to classical oscillators the IG sources may be found outside the allowed region $V$ like in quantum oscillators. 

The analysis of Eq.(\ref{FEx}) in case of a $3D$-wave operator $\nabla^2_{\tau \mathbf{r} } = \frac{\partial^2}{\partial \tau^2} - \nabla^2_{\mathbf{r}} $ is similar to just considered $1D$-case of Eq.(\ref{1W}): 
\begin{equation} 
\frac{\partial^2 \phi (\tau,\mathbf{r})}{\partial \tau ^2} - \nabla^2_{\mathbf{r}} \phi (\tau,\mathbf{r} ) = \sum^N_{n=1} q_n (\tau) \delta (\mathbf{r}_n).  
\label{3W} 
\end{equation} 

Representing as before the IG sources by harmonic oscillators: $q(\tau, \mathbf{r} _n) = q \delta(\mathbf{r} _n) e^{\imath \omega \tau}$, from Eq.(\ref{3W}) one can get fields: 
\begin{equation} 
\phi(\tau, \mathbf{r} ) = \sum^N_{n=1} \phi_n(\tau, \mathbf{r} ) = \sum^N_{n=1} \alpha q_n e^{\imath \omega (\tau - |\mathbf{r} -\mathbf{r} _n|)}, 
\label{3F} 
\end{equation} 
which are a superposition of spherical waves spreading from each point $\mathbf{r}_n$. 

Two harmonic sources distributed on the spheres of radiuses $R_k$ and $R_n$: $q(R_i) = q_i \delta (R_i) e^{\imath \omega \tau} Y^m_l(\vartheta,\varphi)$, where $ R_i=|\mathbf{R}_i|$, $Y^m_l$ are spherical harmonics, can form a stable IG system, if the radiuses $R_k$ and $R_n$ are zeros of the spherical Bessel function of the first kind: $\frac{1}{\sqrt{R_i}} J_{l+\frac{1}{2}}(\omega R_i) =0$. In this case, the standing waves exist in the layer $[R_k,R_n]$ between spheres and waves compensate each other outside it. As far as a center of a sphere is excluded, the spherical Bessel functions of the second kind: $\frac{1}{\sqrt{R_i}} Y_{l+\frac{1}{2}}(\omega R_i) =0$ are also available for the description of the steady-states of IG systems. 

\begin{figure*}[t]
	\begin{center}
		\includegraphics*[angle=270,width=175mm]{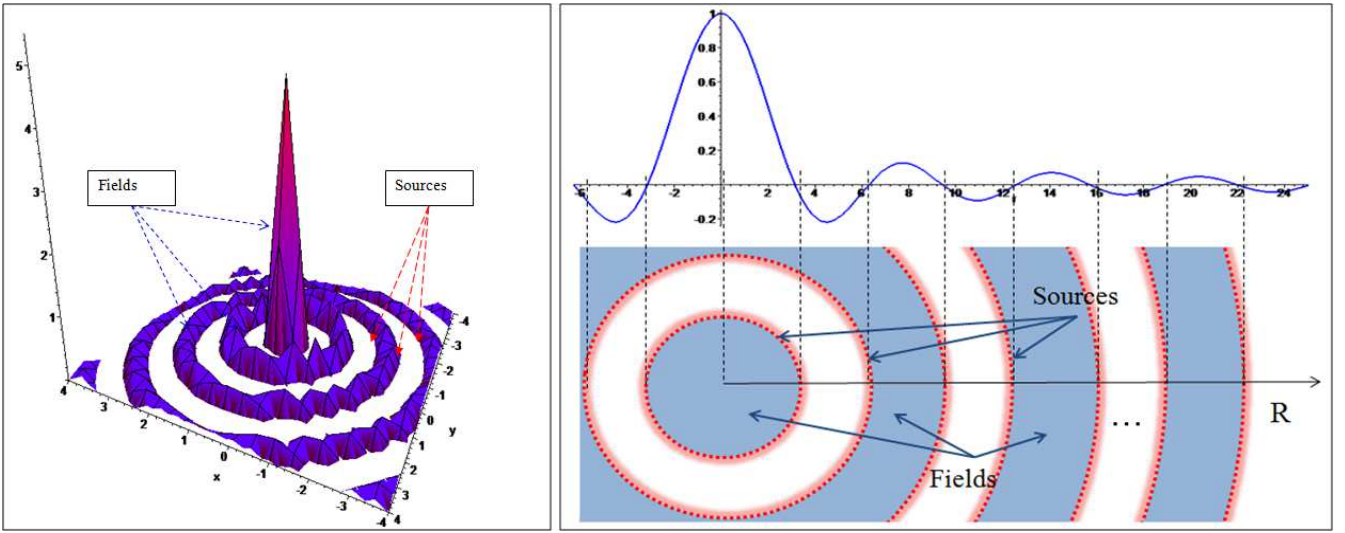}
	\caption{Sources and Fields in $3D$ Spherical IG System} 
	\label{fig5}
	\end{center}
\end{figure*}

The harmonic sources distributed on only one sphere of radius $R_n$ can also form a stable IG system, if the radius $R_n$ are zeros of the spherical Bessel function of the first kind: $\frac{1}{\sqrt{R_n}}J_{l+\frac{1}{2}}(\omega R_n) =0$. Odd zeros correspond to sources in the same phase and even zeros to sources in antiphase to each other at opposite points of the sphere. The standing waves exist inside the sphere and waves compensate each other outside it. Thus, a stable spherically symmetric IG system is a $3D$ sphere with allowed and forbidden radiuses for fields and sources (see Figure~\ref{fig5}). 
 
A dynamics of spherical sources is similar to just considered case of a $1D$-wave operator: the sources oscillate non-linearly near their radiuses of equilibrium. 
An observer interprets the standing waves as if their sources are in permanent exchange by quanta of fields with each other. It correlates with a mechanism of interactions in field theories (this question will be discussed in more details in Section~\ref{sec:PhC}). In passing between steady-states, the IG system emits or absorbs the quanta of fields. 

The considered mechanism of matching of sources and fields is inherent to macro-objects, to so-called fundamental optical solitons, which are the compacts of electromagnetic fields appearing in optical media during the laser pumping (see, for example, N.Rosanov\cite{Ros07}). Solitons behave as separate physical systems, can interact with each other, merge or divide into different parts, possess explicit quantum characteristics. Probably, other mechanisms of the source-field stabilization, for example, in vortical solitons, may also be suitable for IG systems. 

\section{\bf Physical Correspondences}
\label{sec:PhC}

Considering the coupling, we have not specified the parameters of off-site spaces and shared regions: their dimensionality, connectivity, scale, etc. We have used only a stability in time of the coupling (or, more exactly, a quasi-stability, when processes between spaces are much faster then inside them). Due to Eq.(\ref{qD}) the coupling is identified by the observer with physical objects or systems on shared regions. Specifying additional characteristics to shared regions or off-site spaces one may describe quite different physical objects. 

For example, if the shared region $D$ is not a connected domain, but consists of two or more parts: $D$: $D_1 \cup D_2 ... \cup D_N$, so the same physical system may exist `simultaneously' in some spatially separated regions. The effects of Einstein-Podolsky-Rosen of the immediate transfer of some characteristics between `different particles' (such as spin, etc.) just get a clear interpretation.  

Furthermore, one may consider the shared region $V$ in different scales to the observer. At a micro-scale the IG systems may be identified with quantum physical objects, such as elementary particles, nuclei, atoms, the Cooper pairs of electrons in superconductors, etc. At a scale of the observer the similar IG systems are solitons, considered above, or, probably, fire balls. It is cosmological objects: galaxies and the universe at a macro-scale to the observer. 
The bigger scale we consider, the lower stability of objects -- IG systems we have, because micro-objects may destroy completely or partially the matching of fields and sources in macro IG systems. Thus, a scale of the observed objects is defined by the local energy level of the background of the universe. Indeed, particles, nuclei, atoms may be stable, but are destroyed at high temperatures and densities of surrounding media, for example, inside stars or at early `hot' universe. The middle-scale solitons and fire balls are quasi-stable because of the losses in surrounding media. The macro-scale cosmological objects: galaxies and the universe are quasi-stable or even unstable. 

The matching of IG systems at micro scale is identified with the confinement and strong interactions. Similar matching in cosmological IG systems is interpreted by an observer from the containing space as invisible gravitating masses -- a dark matter, while the same matching looks for an observer from the included space as invisible anti-gravitating masses -- a dark energy (see Section~\ref{sec:OSS}). Different influences of dark essences (which lead, for example, to different rotation curves of spiral galaxies) may be explained by different degree of partial matching in quasi-stable cosmological IG systems. 

An energy-momentum four-vector of a stable IG system is: $E^2-p^2 c^2 = m^2 c^4$, where $E$ is the total energy, $p$ a momentum and $m$ its mass. In intrinsic frame of reference: $p=0$, so $E= m c^2$ and an energy of standing waves inside a spatial region $V$ looks like a part of an intrinsic mass of the physical object. An observer may conclude that field quanta of these standing waves have a mass, even if they are massless (for example, photons). It exactly occurs in electroweak theory, where photons $\gamma$ are being generalized with massive $W^\pm$ and $Z^0$ bosons. 

A `charge' of $W^\pm$ bosons may also be `explained' by a rough interpretation of observations. Indeed, if the standing waves are interpreted as an exchange of quanta between sources, so an observer will see that sources change their charges during the quanta exchange. He may conclude that the wave quanta transfer a charge from one source to another, so quanta possess a charge. 

\begin{figure}[t]
	\begin{center}
		\includegraphics*[angle=270,width=85mm]{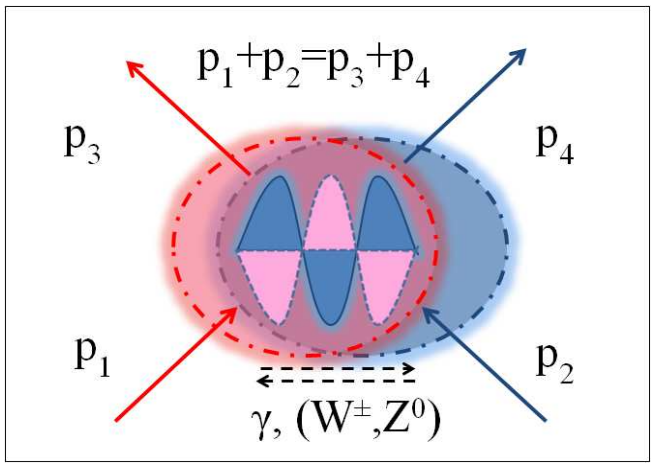}
	\caption{Interaction of particles} 
	\label{fig6}
	\end{center}
\end{figure}

In fact, such `miracles' are just consequences of a rough model of particles' interactions by an exchange of field quanta, that is usually illustrated in field theories by Feynman diagrams. Indeed, an attraction of particles by an exchange of field quanta looked artificial and farfetched from the very beginning. The model of interactions in frame of IG systems looks more natural: the standing waves only look like an exchange of field quanta, but they are not a reason of particles' interactions, they are an integral part of an IG system (see Figure~\ref{fig6}). Thus, particles' interactions, their coalescence and decays are just reorganizations of the IG systems. 

Let's consider the IG systems created by the IG sources $q_\xi (D) = q \delta_\xi (D)$, where $q=0$. There is an infinite number of such systems, because there is an infinite number of sets $\{q_n\}$: $q=\sum^N_{n=1} q_n =0$ with $N\to \infty$. These `zero IG systems' are new physical objects, which do not have physical characteristics (because $q=0$) and do not have fields outside shared regions $D$. An observer identifies such physical objects as `nothing' or as a model of a {\it physical vacuum}. This way, the energy of standing waves in $D$ is  a part of an intrinsic energy of a physical vacuum. 

Clear correspondences exist between stable IG systems and the `Cooper pairs' of electrons in superconductivity. Moving in this direction, one may find analogies with the `all-pervasive' Higgs fields (see, for example, L.Ryder\cite{Ryd09}). In an IG formalism Higgs fields look like some of interpretations of a `multiple content' of IG functions. The `multiple content' of IG functions on regions $D$ or $V$ is a fundamental property, which we will discuss in Section~\ref{sec:CoR}. 

\section{\bf Cardinality of Reality}
\label{sec:CoR}

Some of us still believe that any effect has its explicit reason (we use the word `explicit', because we do not mean that there are no `reasons' at all!) and it is exactly the aim of science to discover deterministic physical laws and to create a deterministic model of the universe. These people believe that it is possible to interconnect by cause-effect chains all events in the world around at any infinitesimal interval in space and time. This paradigm predetermines the use of real numbers (aligned one by one without `gaps') and appears as connectivity and continuality of a spacetime in physical theories. 

However, a mathematical theory of sets\cite{Can874,Cohen66} declares that not all, but only specific events may be considered in frames of a spacetime continuum or a manifold. The power of these events is limited by a transfinite cardinality of continuums $\aleph_1$, but, generally, there are much more powerful sets with cardinalities $\aleph_2$, $\aleph_3$, etc. up to infinity. Do we have any objective reason to limit the transfinite cardinality of events in Reality by $\aleph_1$? Most probably, no. Most probably, the use of explicit causal models and deterministic schemes is defined by a `subjective' reason -- the neurophysiology of a human brain. Of course, deterministic schemes are comfortable for us, give us a possibility to survive, but it does not mean that such explicit causal models are complete. 

There are no objective reasons to limit the cardinality of events in Reality by any level, so, most probably, its cardinality is $\aleph_\infty$. Thus, both our spacetime and off-site spaces may be considered as subsets of all-containing set of extra-high transfinite cardinality (it is shown on Figure~\ref{fig1}) and the couplings $X$ between spaces and explicit deterministic interconnections $F$ inside spaces are just different parts of total interconnections of the physical objects with the rest parts of Reality. 

To argue in favor of these statements, let's consider a countable set of rational numbers ${\mathbb Q}$ of a transfinite cardinality $\aleph_0$ and a set of real numbers ${\mathbb R}$ of $\aleph_1$. From one side, it is possible to approximate any point $x$ on real axis by rational number with any desired accuracy, because $\forall x \in {\mathbb R}$, $\forall \epsilon >0$: $\exists q \in {\mathbb Q}$: $|x-q|<\epsilon$. It seems enough to use only rational numbers instead of real ones. But, from other side, real numbers are much more powerful than rationals, so, for example, any point of real axis chosen accidentally will be real, but not rational, i.e. a possibility to get accidentally a rational number is equal to zero. From the point of view of the observer of real numbers an observer of rational ones perceives practically nothing. Probably, in its turn, an observer of sets of $\aleph_2$ can tell that an observer of a spacetime continuum perceives practically nothing in Reality. 

Moreover, when an observer of rational numbers uses only operators of summation, subtraction, multiplication or division (the usual axiomatic background of linear spaces), he will never reach real numbers, because for $\forall q_1,q_2 \in {\mathbb Q}$: $ q_1\pm q_2, q_1\cdot q_2, q_1/ q_2 \in {\mathbb Q}$, so his representations will be closed and limited only by rational numbers. Remember, how it was difficult to discover irrational numbers at ancient times! 

By definition of real numbers, the real axis is so `dense' that it is impossible to insert any element between two nearest-neighbor reals. It looks like there is no any place for the elements of more powerful sets. The key word here is {\it between}, because it means that real numbers are arranged in some order, so the elements of more powerful sets need to be non-arranged or arranged differently from this real axis. Different arrangements correspond to different causalities. Exactly such differently arranged spaces have been introduced above as spaces off-site to the observer's spacetime. Of course, any continuous function in some space is being perceived like discontinuous in the off-site spaces, exactly as declared before. 

There are persuasive evidences of existence of processes exceeding deterministic ones. It is well known and many times experimentally proved postulates and paradoxes of quantum theories, such as: an uncertainty relation, a particles' duality, a wave function collapse, EPR-like effects, etc. Indeed, nobody can predict the exact place of the electron on the screen during interference or diffraction and one needs to use, for example, probabilistic methods for physical description. In fact, an uncertainty relation shows a limit of explicit deterministic description in a scale of elementary particles. Many years ago Einstein, the founder of relativistic theories, said: \emph{``God does not play dice''}, having in mind methods of quantum mechanics, but later he admited his mistake in a joke: \emph{``As far as the laws of mathematics refer to reality, they are not certain, as far as they are certain, they do not refer to reality''}.  

The methods of quantum physics including probabilistic ones, fractals, $p$-adic, extended reals, etc. (see, for example, E.E.Rosinger\cite{Ros09} and references in it) may be considered as efforts to describe processes more powerful than $\aleph_1$. Note, that, for example, physical theories with additional hidden dimensions in a spacetime cannot overcome deterministic schemes, because a cardinality of any multi-dimensional continuum or manifold is still $\aleph_1$. 

The IG functions introduced in Definitions~(\ref{qDD},\ref{dD}) are not a point-like functions, but are an integral characteristic of regions $D$ or $V$. Thus, IG functions have a `multiple content' on these regions, i.e. take into account many functions at once, including discontinuous and non-integrable. A cardinality of such `multiple content' is $\aleph_2$ or higher, so an uncertainty of IG functions has a fundamental character and they cannot be reduced to definite combinations of continuous functions.

Such nondeterministic properties of objects of quantum physics as a particles' duality, a wave function collapse, EPR-like effects, etc. exceed the deterministic representations of relativistic theories and look like paradoxes for them. Differences between relativistic and quantum approaches become quite obvious in an IG formalism: objects of quantum physics are off-site to the observer's spacetime and so off-site for his causality. Therefore, relativistic and quantum theories have different causalities, and their backgrounds are and need to be inconsistent with each other.

An IG formalism reveals a very close connection between an identification of physical objects by the observer and their description. In fact, in relativistic theories the observer interprets the Reality from his spacetime, so from his `humancentric' point of view. The `multicentric', multi-space scheme of the IG formalism can include consistently different `points of view', different causalities, different `backgrounds', so, in particular, can `unify' relativistic and quantum theories. 

Ancient philosopher Plato considered that \emph{`the man has the only possibility to see the distorted shadows from the bright and multicolor Reality on the curved wall of the cave, where he is confined and chained up back to the entrance'}. Indeed, we are `captured' by explicit cause-effect chains inside a `cave' of our logic and can only perceive the multi-spaces' `multicolor Reality' through `distorted shadows' of the `curved wall' of our spacetime representations. Probably, determinism is a necessary way of human cognition, but, certainly, not a limit. 

\section*{\bf Conclusion}

There are no reasons to deny an existence of many `universes' appeared by the same or similar way as our universe. Also, nobody has objective reasons to limit the cardinality of events in the Reality, so our spacetime and off-site spaces may be considered as subsets of extra-powerful Reality with the transfinite cardinality of events $\aleph_\infty$. Explicit deterministic schemes built in frames of the spacetime continuum or manifold cannot exceed the cardinality $\aleph_1$, so they are incomplete. The use of the multi-space structure of the universe extends the cardinality of events available for physical description. 

The coupling of off-site spaces is being identified by the observer with physical objects or systems. Deterministic interconnections between physical objects inside a space and nondeterministic interconnections with off-site spaces are parts of general interconnections of physical objects in quite powerful Reality. Specifying characteristics of shared regions or off-site spaces, one may describe quite different physical objects: from elementary particles, nuclei, atoms at a micro-scale to the observer up to cosmological objects: galaxies and the universe at a macro-scale. 

Experimentally proved paradoxical properties of quantum physical objects and interactions, the existence of dark matter and dark energy in modern cosmology are persuasive proofs of an existence of nondeterministic processes and physical objects. Both deterministic and nondeterministic descriptions of physical objects are unified in the IG systems. A model of particles' interactions by an exchange of field quanta in field theories seems rough. In an IG formalism the model of interactions looks more natural: interactions of particles, their coalescence and decays are just reorganizations of the IG systems.

Relativistic theories are deterministic schemes built in frames of a spacetime continuum or manifold, while quantum theories describe nondeterministic properties of physical objects defined by the coupling with off-site spaces, so foundations of relativistic and quantum theories are and need to be inconsistent with each other. A multi-space model of the Universe is a multi-causality scheme, so relativistic and quantum theories may be consistently `unified' in it. Determinism seems necessary, but not the only way of human cognition.

\end{document}